\pdfoutput=1
\documentclass[%
 reprint,
 amsmath,amssymb,
 aps,
pra,
]{revtex4-1}

\usepackage{graphicx}
\usepackage{dcolumn}
\usepackage{bm}
\usepackage{hyperref}

\begin{document}

\preprint{APS/123-QED}

\title{The Singular Optical Axes in Biaxial Crystals and\\
Analysis of Their Spectral Dispersion Effects in $\beta$-Ga$_2$O$_3$}

\author{Marius Grundmann}
 \email{grundmann@physik.uni-leipzig.de}
\author{Chris Sturm}%
 \email{csturm@physik.uni-leipzig.de}
\affiliation{%
 Universit\"at Leipzig, Institut f\"ur Experimentelle Physik II, Linn\'estr. 5, D-04103 Leipzig, Germany}%


\date{\today}

\begin{abstract}
We classify and distinguish optically biaxial materials, which can have triclinic, monoclinic or orthorhombic crystal symmetry, by the degeneracy of the indices of refraction of their four singular optical axes ({\it Windungsachsen}) in the absorption regime. We provide explicit analytical solutions for angular orientations of the singular optical axes in monoclinic crystals and orthorhombic crystals.
As a model material we analyze monoclinic gallia ($\beta$-Ga$_2$O$_3$) and discuss in detail the dispersion (i.e. the
spectral variation of the angular position) of its singular optical axes. For a certain energy range ($E \approx 7.23$--$7.33$\,eV) we find quasi-uniaxial symmetry.  At two energies ($E \approx 8.14$\,eV and $E \approx 8.37$\,eV) we find triaxial spectral points for which one regular optical axis and two singular optical axes exist. Concurrently a Stokes analysis of the spectral dependence of the electrical field eigenvectors is made and discussed for various crystal orientations. For a singular optical axis $|S_3|=1$; for the two degenerate singular axes at the triaxial point the Stokes vector is undefined. For a certain energy ($E=6.59$\,eV), the $\langle 010 \rangle$-orientation is close to a singular optical axis, $|S_3|=0.977$. The analysis provided here is prototypical for the treatment of the optical properties of optically biaxial functional materials in the absorption and gain regimes.
\end{abstract}

\pacs{xx.xxxx}
\maketitle


\section{Introduction}

It was pointed out first by Voigt in 1902~\cite{voigt} that the two optical axes, i.e. crystal
directions in which the index of refraction does not depend on polarization,
of biaxial crystals split into four \textit{Windungsachsen}, now mostly termed 'singular optical axes',
in the absorption regime when the elements of the dielectric tensor are complex.
Three crystal systems are optically biaxial, namely, in increasing order of symmetry,
the triclinic, monoclinic and orthorhombic systems. In~\cite{voigt} the singular optical
axes were discussed for an orthorhombic crystal assuming
the imaginary part of the dielectric tensor to be small. The complete problem for
orthorhombic symmetry was elaborated in~\cite{berek}. In recent literature the general case,
also including chiral contributions has been discussed theoretically~\cite{berry}. 

The forms
of the dielectric tensor of triclinic ($\epsilon_{\mathrm{t}}$), monoclinic ($\epsilon_{\mathrm{m}}$, for the non-right angle being in the $(x,z)$-plane)
and orthorhombic ($\epsilon_{\mathrm{o}}$) crystals without chirality
are symmetric and given by
\begin{subequations}
\begin{eqnarray}
\epsilon_{\mathrm{t}} &=& \left( \begin{array}{ccc} \epsilon_{xx} & \epsilon_{xy} & \epsilon_{xz} \\ 
\epsilon_{xy} & \epsilon_{yy} & \epsilon_{yz} \\ \epsilon_{xz} & \epsilon_{yz} & \epsilon_{zz} \end{array} \right) \;, \\
\epsilon_{\mathrm{m}} &=& \left( \begin{array}{ccc} \epsilon_{xx} & 0 & \epsilon_{xz} \\ 
0 & \epsilon_{yy} & 0 \\ \epsilon_{xz} & 0 & \epsilon_{zz} \end{array} \right)  \;, \\
\epsilon_{\mathrm{o}} &=& \left( \begin{array}{ccc} \epsilon_{xx} & 0 & 0\\ 
0 & \epsilon_{yy} & 0 \\ 0 & 0 & \epsilon_{zz} \end{array} \right) \label{eq:eps_o}\;.
\end{eqnarray}
\end{subequations}
In the transparency regime the tensor elements are real ($\epsilon=\epsilon'$), in the absorption regime they are complex ($\epsilon=\epsilon'+\imath \, \epsilon''$).

We note here that it is trivial and well known that for uniaxial crystals, e.g. $\epsilon_{xx}=\epsilon_{yy}$ in eq. (\ref{eq:eps_o}), there is a single optical axis regardless whether $\epsilon$ is real or complex.

The discussion of the energy dependence of the direction of the singular optical axes for real biaxial materials 
has not been reported in the literature since spectrally resolved data for the dielectric tensor of such materials are rarely reported. Here, we first give a general discussion of the singular optical axes and the degeneration of their indices 
of refraction for the triclinic, monoclinic and orthorhombic systems. 
We also develop an analytical formula
for the angular azimuthal position of the singular axes in monoclinic crystals. It can be simplified for
$\epsilon_{xz}=0$ and provides a complete explicit analytical solution for the orthorhombic case.
Then we use the recently published
complete dielectric tensor for monoclinic $\beta$-Ga$_2$O$_3$ (available in a wide transparency regime 0.5--4.7\,eV
and within the absorption regime from about 4.7\,eV up to 8.5\,eV)~\cite{sturm}. 

The analysis provided here in general and for the model material $\beta$-Ga$_2$O$_3$ is prototypical for the treatment of the optical properties of further optically biaxial functional materials, e.g. used in photodetectors (absorption regime) or lasers (gain regime), both operating in the $\epsilon'' \ne 0$ regime.

\section{Theory}

\begin{figure}[htb!]
\centering
\includegraphics[width=0.5\linewidth]{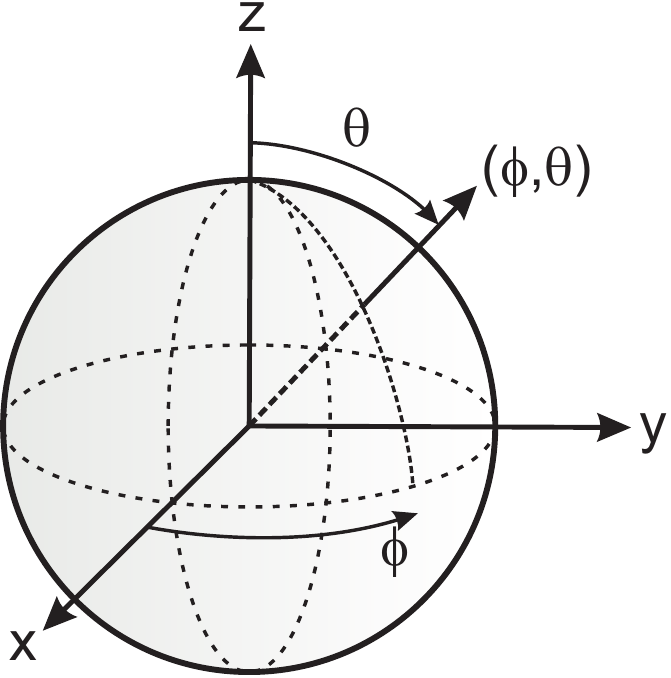}
\caption{\label{fig:sketch}Geometry of rotation operation on the crystal.} 
\end{figure}

We assume an incoming electromagnetic wave with wave vector along the $z$ direction.
The dielectric tensor is rotated by the polar angle $\theta$ around the $y$-axis
and subsequently by the azimuthal angle $\phi$ around the $z$-axis (Fig.~\ref{fig:sketch}), 
\begin{equation}
R=
R_z(\phi) \, R_y(\theta) = 
\left(
\begin{array}{ccc}
 \cos \theta \, \cos \phi  & -\sin \phi  & \cos \phi \, \sin \theta  \\
 \cos \theta \, \sin \phi  & \cos \phi  & \sin \theta \, \sin \phi  \\
 -\sin \theta  & 0 & \cos \theta  \\
\end{array}
\right) \;.
\end{equation}
We note that all algebraic and numerical calculations have been performed with
Mathematica~\cite{mathe}.

%

The problem of finding the optical axes can be formulated in the electric field $\mathbf{E}$~\cite{yeh,emslie},
\begin{equation}
\left| R^{-1} \, \epsilon \, R - n^2 \, \left( \begin{array}{ccc} 1 & 0 & 0 \\ 0 & 1 & 0 \\ 0 & 0 & 0 \end{array} \right) \right|=0 \;.
\label{eq:det_E}
\end{equation}
$n$ denotes the generally \textit{complex} index of refraction that has an imaginary part in the absorption regime. With $\mathbf{E}=\epsilon^{-1} \mathbf{D}$ in mind and
multiplying (\ref{eq:det_E}) with $R^{-1} \, \epsilon^{-1} \, R$, the similar equation in the displacement field $\mathbf{D}$ is~\cite{berry},
\begin{equation}
\left| R^{-1} \, \epsilon^{-1} \, R \, \left( \begin{array}{ccc} 1 & 0 & 0 \\ 0 & 1 & 0 \\ 0 & 0 & 0 \end{array} \right) - \frac{1}{n^2} 
\left( \begin{array}{ccc} 1 & 0 & 0 \\ 0 & 1 & 0 \\ 0 & 0 & 1 \end{array} \,  \right) \right| =0 \;.
\label{eq:det_D}
\end{equation}
The latter approach uses nicely the transversality of the $\mathbf{D}$-field. One of the three eigenvalues of (\ref{eq:det_D}) is zero  and the equation becomes an 
eigenvalue problem with a complex symmetric $2 \times 2$ matrix,
\begin{equation}
\left| \, \left( \begin{array}{cc} [R^{-1} \, \epsilon^{-1} \, R]_{xx} & [R^{-1} \, \epsilon^{-1} \, R]_{xy} 
\\ {[R^{-1} \, \epsilon^{-1} \, R]_{xy}} & {[R^{-1} \, \epsilon^{-1} \, R]_{yy}} \end{array} \right) - \frac{1}{n^2} 
\, \left( \begin{array}{cc} 1 & 0  \\ 0 & 1  \end{array} \,  \right) \right| =0 \;.
\label{eq:hsd9whd}
\end{equation}

The orientations $(\phi,\theta)$ of the optical axes are found when the two (complex) solutions $n_1$ and $n_2$ of (\ref{eq:det_E}) or equivalently of (\ref{eq:det_D}) are the same.

There are generally eight solutions for biaxial crystals for the parameter range $0 \le \phi \le 2 \pi$
and $0 \le \theta \le \pi$, related to the four singular axes and their forward and backward intersections with the unity sphere. Along the (singular) optical axes in the dissipative regime, only one of the two circularly
polarized waves can propagate, thus Voigt called these optical axes \textit{Windungsachsen}~\cite{voigt}. The mathematical reason
is that at these singular axes the two eigenvectors of the matrix problem (\ref{eq:det_D}) are collinear and span
only a one-dimensional space, leading to so-called Voigt waves~\cite{mackay,gerardin}. This is a general property of complex symmetric matrices~\cite{heiss}.

The two solutions $n_{1,2}$ determined from (\ref{eq:det_E}) are ($\tilde \epsilon=R^{-1} \, \epsilon \, R$)
\begin{equation}
n^2_{1,2}=\frac{(\tilde \epsilon_{xx} + \tilde \epsilon_{yy}) \, \tilde \epsilon_{zz} - \tilde \epsilon_{xz}^2 - \tilde \epsilon_{yz}^2  \pm \sqrt{r}}{2 \, \tilde \epsilon_{zz}} \;.
\end{equation}
with $r(\epsilon,\phi,\theta)$ given by
\begin{eqnarray}
r=[(\tilde \epsilon_{xx} + \tilde \epsilon_{yy}) \tilde \epsilon_{zz}-\tilde \epsilon_{xz}^2 - \tilde \epsilon_{yz}^2]^2 + \\
4 \tilde \epsilon_{zz} [\tilde \epsilon_{xz}^2 \tilde \epsilon_{yy} - 2 \tilde \epsilon_{xy} \tilde \epsilon_{xz} \tilde \epsilon_{yz} + \\
\tilde \epsilon_{xy}^2 \tilde \epsilon_{zz} + \tilde \epsilon_{xx} (\tilde \epsilon_{yz}^2 - \tilde \epsilon_{yy} \tilde \epsilon_{zz})] \;.
\label{eq:cbweu4}
\end{eqnarray}
The requirement of $n_1=n_2$ for an optical axis leads to the condition
\begin{equation}
\Delta=(n_1^2-n_2^2)^2=\frac{r}{\tilde \epsilon_{zz}^2}=0 \;,
\label{eq:cud8w4}
\end{equation}
or just $r=0$. Simpler analytical expressions for $r$ actually
follow from the equivalent equation to (\ref{eq:det_E})
\begin{equation}
\left| \, \epsilon \, - \, n' \, ^2 \, R \, \left( \begin{array}{ccc} 1 & 0 & 0 \\ 0 & 1 & 0 \\ 0 & 0 & 0 \end{array} \right) \, R^{-1} \right| =0 \;,
\end{equation}
leading to $r'=0$, with
\begin{widetext}
\begin{eqnarray}
r'_m&=&[\cos ^2 \!\phi \, \left(\epsilon_{xx} \epsilon_{zz}-\epsilon_{xz}^2+\epsilon_{yy} \epsilon_{zz} \cos ^2 \!\theta  \right)+\sin ^2 \!\phi \, \left(\cos ^2 \!\theta \, \left(\epsilon_{xx} \epsilon_{zz}-\epsilon_{xz}^2 \right)+\epsilon_{yy} \epsilon_{zz} \right) 
+\epsilon_{xx} \epsilon_{yy} \sin ^2 \!\theta  +\epsilon_{xz} \epsilon_{yy} \sin 2 \theta \, \cos  \phi ]^2 \nonumber \\
&& + 4 \epsilon_{yy} [\epsilon_{xz}^2-\epsilon_{xx} \epsilon_{zz} ] [ \epsilon_{xx} \sin ^2 \!\theta \, \cos ^2 \!\phi  
+ \epsilon_{xz} \sin 2 \theta \, \cos^3 \!\phi + \epsilon_{xz} \sin  2 \theta \, \sin ^2 \!\phi \, \cos \phi +
\epsilon_{yy} \sin^2 \!\theta \, \sin ^2 \!\phi + \epsilon_{zz} \cos^2 \!\theta ]
\label{eq:rs_m} \\
r'_o&=&[\epsilon_{zz} \cos ^2\! \phi \, \left(\epsilon_{xx}+\epsilon_{yy} \cos ^2 \!\theta \right)+\epsilon_{zz} \sin ^2 \!\phi \, \left(\epsilon_{xx} \cos ^2 \!\theta +\epsilon_{yy}\right)+\epsilon_{xx} \epsilon_{yy} \sin ^2 \!\theta ]^2
\nonumber \\
&& -4 \epsilon_{xx} \epsilon_{yy} \epsilon_{zz}  [ \sin ^2 \!\theta \, \left(\epsilon_{xx} \cos ^2 \!\phi +\epsilon_{yy} \sin ^2 \!\phi \right)+\epsilon_{zz} \cos ^2 \!\theta ] 
\label{eq:rs_o} 
\end{eqnarray}
\end{widetext}
for the monoclinic and orthorhombic case. The triclinic case leads to a more lengthy expression and will be treated in detail in a subsequent publication. 

For triclinic material, the values of $n_{1,2}$ (and $r$ or $r'$) at opposite sides of the sphere are identical,
\begin{equation}
n_{1,2}(\phi,\theta) = n_{1,2}(\phi+\pi,\pi-\theta) \label{eq:r2} \;,
\end{equation}
and the indices of refraction of the four singular axes are generally all different from each other.
For a monoclinic crystal the following additional symmetry exists,
\begin{equation}
n_{1,2}(\phi,\theta) = n_{1,2}(-\phi,\theta) = n_{1,2}(2\pi-\phi,\theta) \;,
\label{eq:r1} 
\end{equation}
thus forcing pairwise identical indices of refraction for the singular axes. Two independent solutions for $\phi$ exist.
For an orthorhombic crystal a further symmetry (additionally to (\ref{eq:r2}) and (\ref{eq:r1})) exists,
\begin{equation}
n_{1,2}(\phi,\theta) = n_{1,2}(\phi,\pi-\theta) \;,
\label{eq:r0}
\end{equation}
forcing all singular axes to have the same index of refraction; only one independent solution for $\phi$ exists.

A summary for all crystal systems is given in Table~\ref{tab:4axis} (the case $\epsilon''=0$ is well known). Thus the three types of biaxial crystals in the absorption regime are \textit{fundamentally different by the number of different complex indices of refraction of their singular optical axes}.

\begin{table}[htb!]
  \caption{\label{tab:4axis} Number $N_a$ of optical axes for various crystal symmetries in the transparency regime ($\epsilon''=0$)
  and the absorption/gain regime ($\epsilon'' \ne 0$) (assuming $\epsilon' \ne 0$) and number $N_n$ of different (complex) indices of refraction for these axes}
  \begin{ruledtabular}
  \begin{tabular}{@{}l | ll | ll @{}}
  crystal & \multicolumn{2}{c|}{$\epsilon''=0$} & \multicolumn{2}{c}{$\epsilon'' \ne 0$} \\
  & $N_a$ & $N_n$ & $N_a$ & $N_n$ \\
    \hline
triclinic     & 2 & 1 & 4 & 4 \\
monoclinic    & 2 & 1 & 4 & 2 \\
orthorhombic  & 2 & 1 & 4 & 1 \\
tetragonal    & 1 & 1 & 1 & 1 \\
trigonal      & 1 & 1 & 1 & 1 \\
hexagonal     & 1 & 1 & 1 & 1 \\
cubic         & $\infty$ & 1 & $\infty$ & 1\\
  \end{tabular}
  \end{ruledtabular}
\end{table}

The optical axes are found from the solution of $r=r'+ \imath \, r''=0$, i.e. simultaneously $r'=0$ and $r''=0$. The solutions $(\phi,\theta)$ can be found numerically from the intersections of the curves $r'=0$ and $r''=0$ in the $(\phi,\theta)$-plane. For monoclinic crystals, after a few algebraic transformations, the two independent solutions for $\phi$ are given by
\begin{equation}
\phi_{1,2} = \Re \left[ \arctan \sqrt{\frac{a \pm 2\,\sqrt{b}}{c}} \,\, \right] \;,
\label{eq:dbhjer}
\end{equation}
with
\begin{subequations}
\begin{eqnarray}
a &=& \epsilon_{xz}^2 \, \epsilon_{yy} \, (-\epsilon_{xx} - 2 \epsilon_{yy} + \epsilon_{zz}) \nonumber \\
 && + \epsilon_{xx} \, \epsilon_{yy} \, (\epsilon_{xx} - \epsilon_{zz}) \, (-\epsilon_{yy} + \epsilon_{zz}) \\
b &=& - \epsilon_{xz}^2 \, \epsilon_{yy}^2 \, [\epsilon_{xz}^2 + (\epsilon_{xx} - \epsilon_{yy}) \, (\epsilon_{yy} - \epsilon_{zz})] \nonumber \\
&& \times (\epsilon_{xz}^2 - \epsilon_{xx} \, \epsilon_{zz}) \label{eq:dbhjer55} \\
c &=& \epsilon_{yy}^2 \, [4 \, \epsilon_{xz}^2 + (\epsilon_{xx} - \epsilon_{zz})^2] \;.
\end{eqnarray}
\end{subequations}

We note that for orthorhombic crystals ($\epsilon_{xz}=0$, i.e. $b=0$ in (\ref{eq:dbhjer55})) the solution simplifies to $\phi_{\mathrm{o}}=\phi_1=\phi_2$,
\begin{equation}
\phi_{\mathrm{o}} = \Re \left[ \arctan \sqrt{\frac{\epsilon_{xx} \, (\epsilon_{yy} - \epsilon_{zz})}{\epsilon_{yy} \, (\epsilon_{xx} - \epsilon_{zz})}} \,\, \right] \;,
\label{eq:ckjfz45}
\end{equation}
Considering (\ref{eq:r2}) and (\ref{eq:r1}), this solutions creates four azimuthal positions of the singular axes. For each of them two $\theta$-values exist,
\begin{equation}
\theta_{1,2}=\arccos \left[ \pm \, \frac{a+2 b}{c} \right] \;.
\label{eq:hfgd64s44sa}
\end{equation}
with
\begin{subequations}
\begin{eqnarray}
a &=& \imath \, \epsilon_{zz} \, (\epsilon_{xx}-\epsilon_{yy}) \, \sin 2 \phi \\
b &=& \epsilon_{xx} \, \epsilon_{yy} \, (\epsilon_{xx}-\epsilon_{zz}) \, (\epsilon_{yy}-\epsilon_{zz}) \\
c &=& 2 \, \epsilon_{xx} \, \epsilon_{yy} - \epsilon_{xx} \, \epsilon_{zz} - \epsilon_{yy} \, \epsilon_{zz} \nonumber \\
&& + (\epsilon_{xx} - \epsilon_{yy}) \, \epsilon_{zz} \, \cos 2 \phi \;.
\end{eqnarray}
\end{subequations}
We note that for the solutions $\phi$ of (\ref{eq:ckjfz45}), the argument of the $\arccos$-function is real.

\section{Application to G\lowercase{a}$_2$O$_3$}

\begin{figure}[htb!]
\centering
\includegraphics[width=0.7\linewidth]{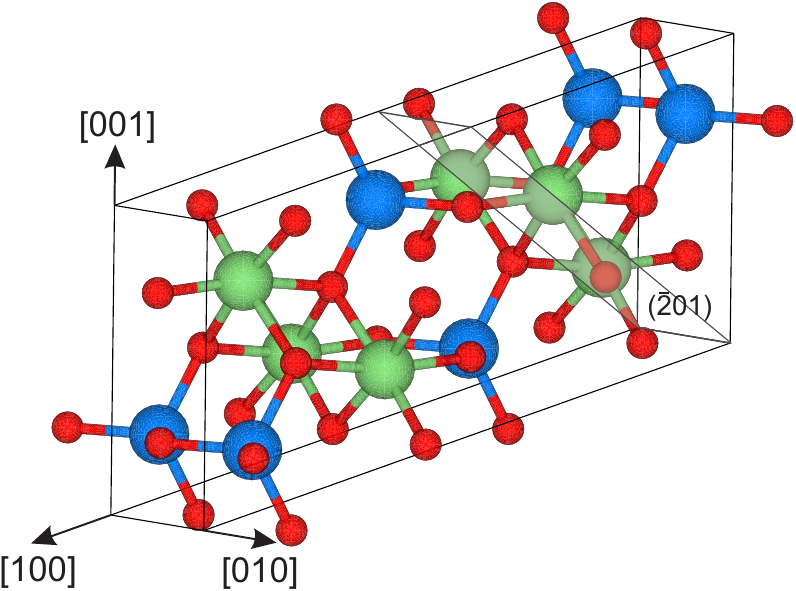}
\caption{\label{fig:sketch2}Unit cell of $\beta$-Ga$_2$O$_3$ (oxygen atoms are shown in \textit{red}, gallium atoms as \textit{green} (octahedral-like coordination) and \textit{blue} (tetrahedral-like coordination)) and its orientation relative to the coordinate system of Fig.~\ref{fig:sketch}:
$[100] \,|| \, x$, $[010] \, || \, y$; the angle between $z$ and $[001]$ is 103.7$^{\circ}-\pi/2$~\cite{kohn}. The ($\bar 2 0 1$) plane
is indicated.} 
\end{figure}

After the proposal of a rather general categorization of biaxial crystals and analytical formulae for the angular positions of the singular optical axes we now turn to the discussion of a real material and some novel effects regarding singular optical axes. For monoclinic
gallia ($\beta$-Ga$_2$O$_3$; a unit cell is depicted in Fig.~\ref{fig:sketch2}~\cite{vesta}) we evaluate the energy dependent complex dielectric tensor reported in~\cite{sturm}. We interpolate the experimental data $\epsilon(E)$ linearly and work 
here with energy steps of 0.01\,eV.
We focus on the optical axes in the absorption regime which starts at about 4.7\,eV. The high energy cut-off of the data is limited by the spectral range of the ellipsometer used in~\cite{sturm}.

\begin{video}[h!]
\href{http://home.uni-leipzig.de/grundm/video_delta_final.gif}
   {\includegraphics[width=0.8\linewidth]{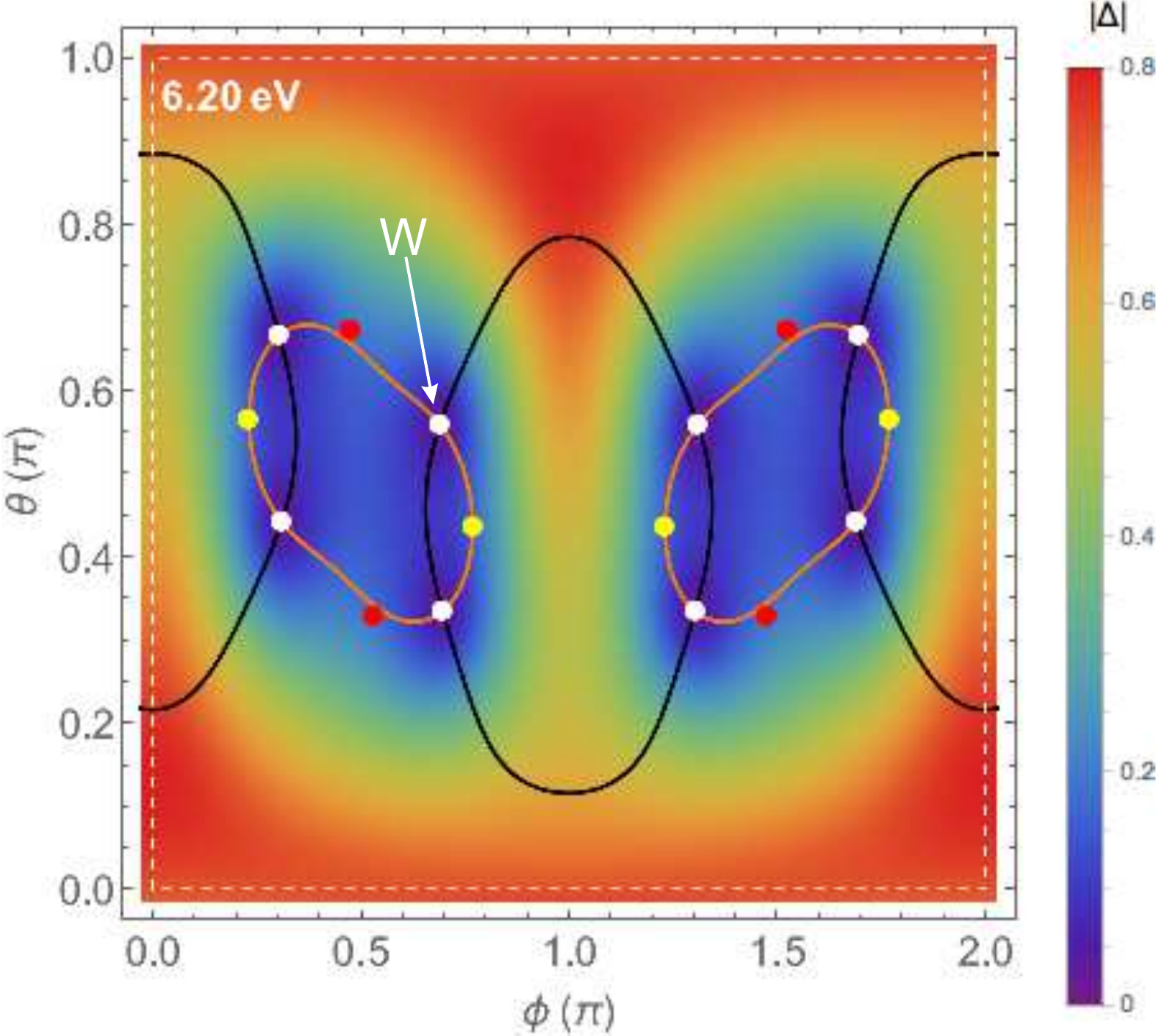}}
 \setfloatlink{http://http://home.uni-leipzig.de/grundm/video_delta_final.gif}
 \caption{\label{vid:dens}%
  The function $|\Delta|$ according to (\ref{eq:cud8w4}) for $\beta$-Ga$_2$O$_3$ (the icon is for $E=6.20$\,eV) in false colors. The curves
  $r'=0$ and $r''=0$ are shown in \textit{black} and \textit{red}. Their intersections ($r=0$), the
  angular positions of the singular optical axes, are marked with \textit{white dots}. The \textit{yellow} and 
  \textit{red dots} mark the positions of the two optical axes of $\epsilon'$ and $\epsilon''$. The
  photon energy is labeled in the upper left corner. The \textit{dashed white rectangle} indicates the
  $(\phi,\theta)$-area for $0 \le \phi \le 2\pi$ and $0 \le \theta \le \pi$. The label 'W' indicates a particular singular 
  optical axis and relates to Fig.~\ref{fig:stokes}.}%
\end{video}
\begin{video}[h!]
\href{http://home.uni-leipzig.de/grundm/video_s3_final.gif}
   {\includegraphics[width=0.8\linewidth]{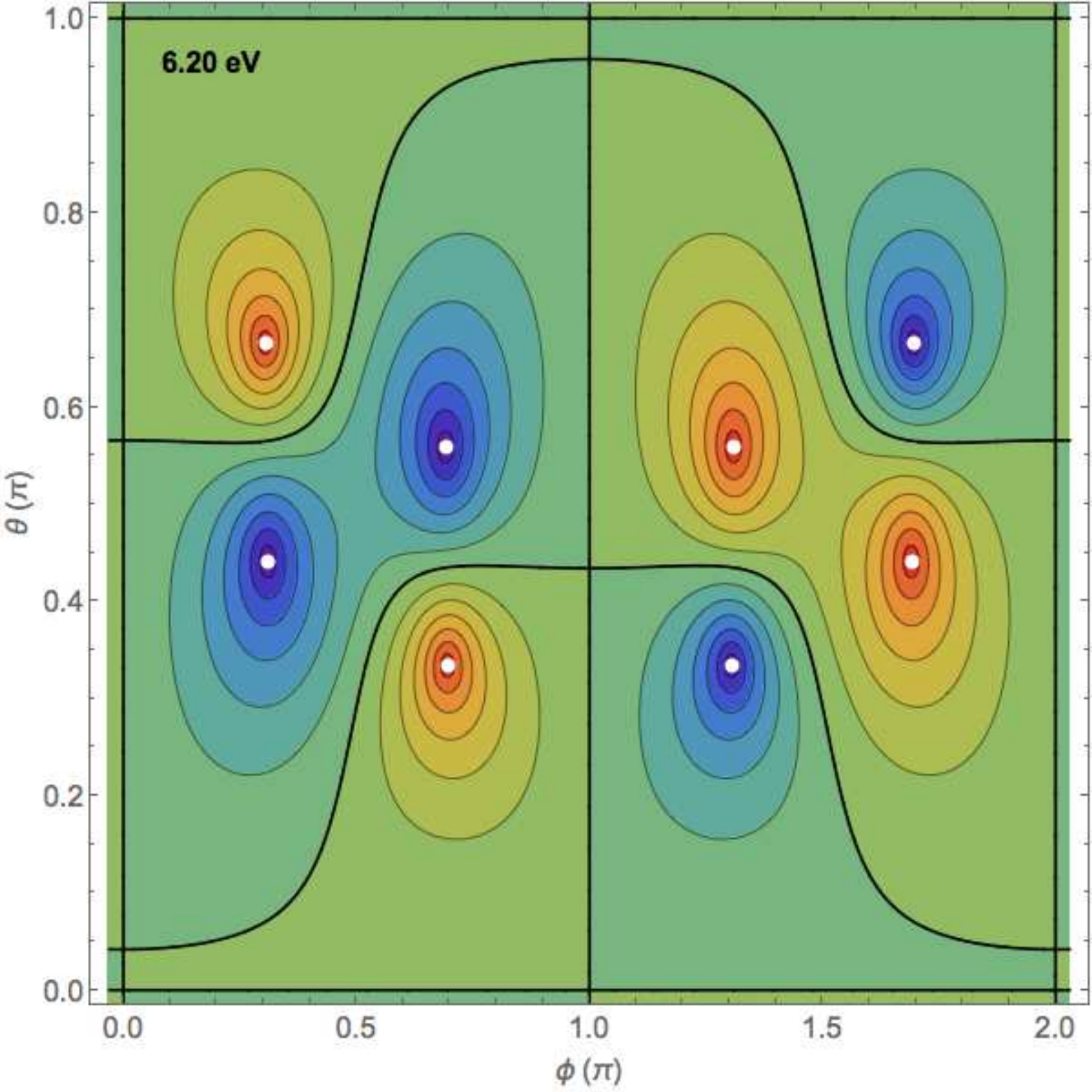}}
 \setfloatlink{http://http://home.uni-leipzig.de/grundm/video_s3_final.gif}
 \caption{\label{vid:s3}%
  The function $S_3$ according to (\ref{eq:S3E3}) for $\beta$-Ga$_2$O$_3$ (the icon is for $E=6.20$\,eV) in false colors (The range $-1 \le S_3 \le +1$ is split with 15 equidistant contour lines with difference $\Delta{S_3}=1/7$; \textit{red}/\textit{blue} indicates	negative/positive values). The
  angular positions of the singular optical axes are marked with \textit{white dots}. The contour with $S_3=0$ (linear polarization) is shown as \textit{thick black lines}. The
  photon energy is labeled in the upper left corner.}%
\end{video}

Video~\ref{vid:dens} illustrates the solution of equation (\ref{eq:cud8w4}) for $\beta$-Ga$_2$O$_3$ at
various energies. The angular positions of the singular optical axes are indicated by white dots.

From the solution (displacement field) eigenvectors $\mathbf{D}_1$ or $\mathbf{D}_2$ of (\ref{eq:det_D}) (at a given ($\phi,\theta$) position) we construct the $\mathbf{E}$-fields ($\mathbf{E}=R^{-1} \, \epsilon^{-1} \, R \, \mathbf{D}$) and their Stokes vectors with the components
\begin{eqnarray}
S_1 &=& E_x \, E_x^* - E_y \, E_y^* \label{eq:S3E1} \;, \\
S_2 &=& E_x \, E_y^* + E_y \, E_x^* \label{eq:S3E2} \;, \\
S_3 &=& - \imath \, ( E_x \, E_y^* - E_y \, E_x^* ) \label{eq:S3E3}\;.
\end{eqnarray}
In the following we use the normalized components, i.e. $S_1^2+S_2^2+S_3^2=1$.
$S_3=\pm 1$ indicates complete circular polarization which occurs for the singular optical axes and Voigt waves. We note that the opposite points of the same singular optical axis have of course opposite sign of $S_3$.
The spectral and angular dependency of $S_3$ is depicted in Video~\ref{vid:s3} for $\beta$-Ga$_2$O$_3$. 
The centers of the circularly polarized regions are the singular optical axes whose positions are indicated by white dots.
The positions with $S_3=0$ (linear polarization) are indicated as thick black lines. 

\begin{figure*}
\centering
(a)\includegraphics[width=0.43\linewidth]{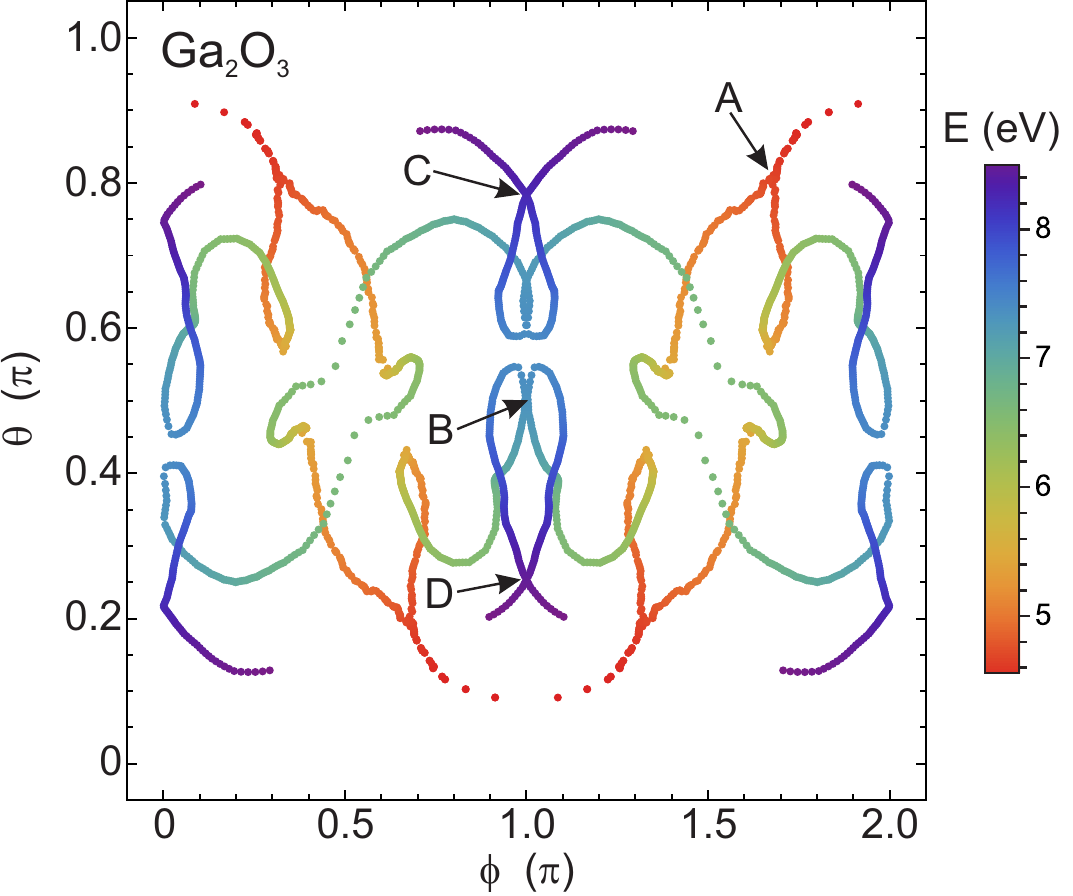}
(b)\includegraphics[width=0.43\linewidth]{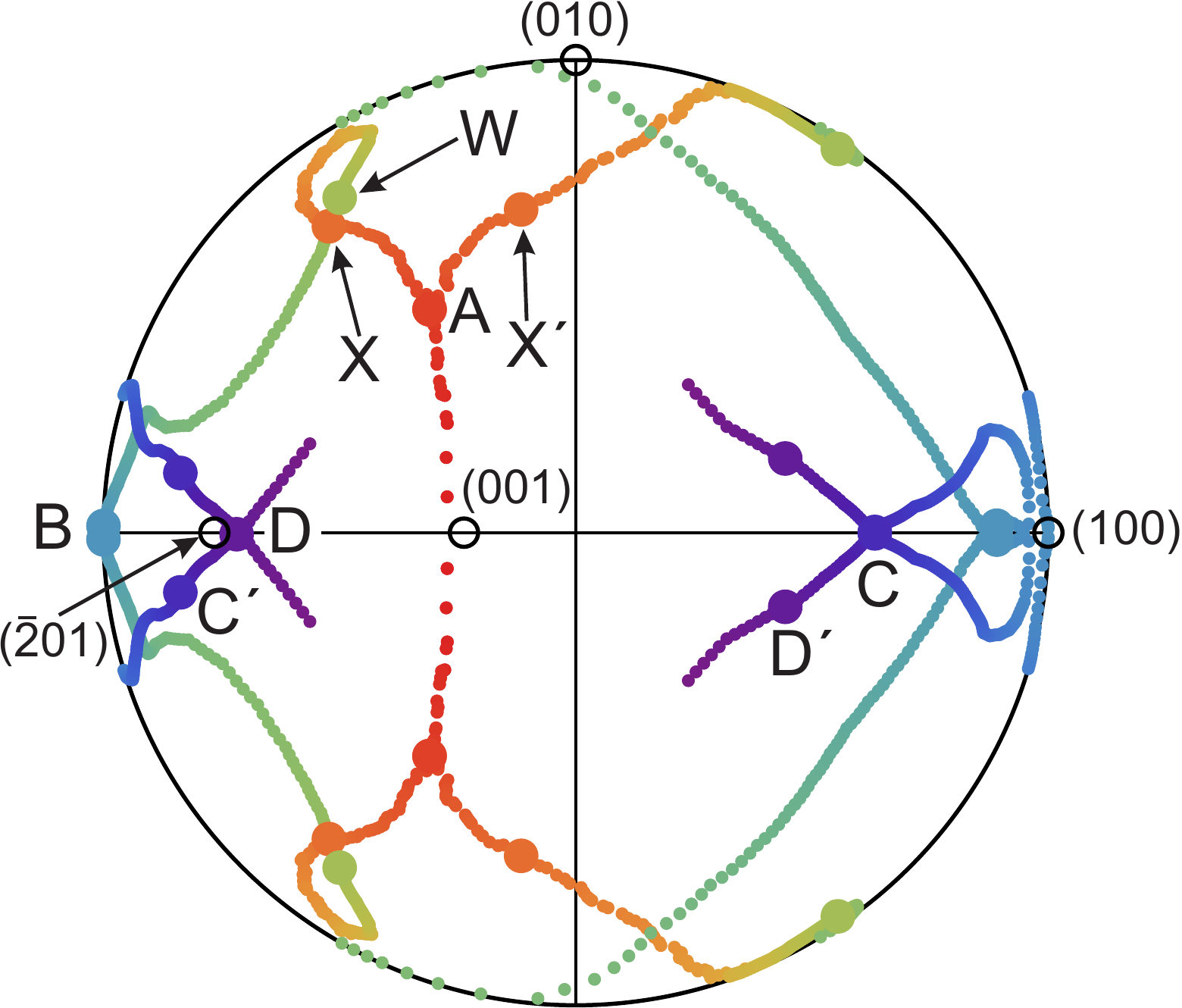}\\
(c)\includegraphics[width=0.43\linewidth]{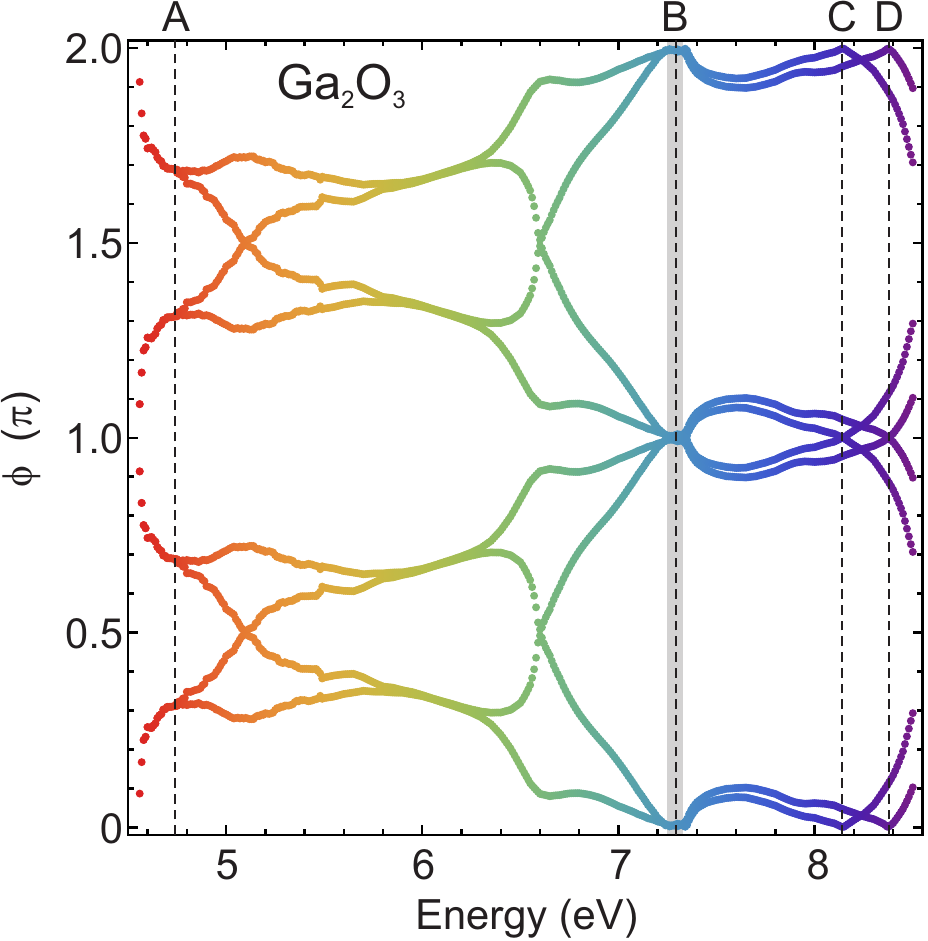}
(d)\includegraphics[width=0.43\linewidth]{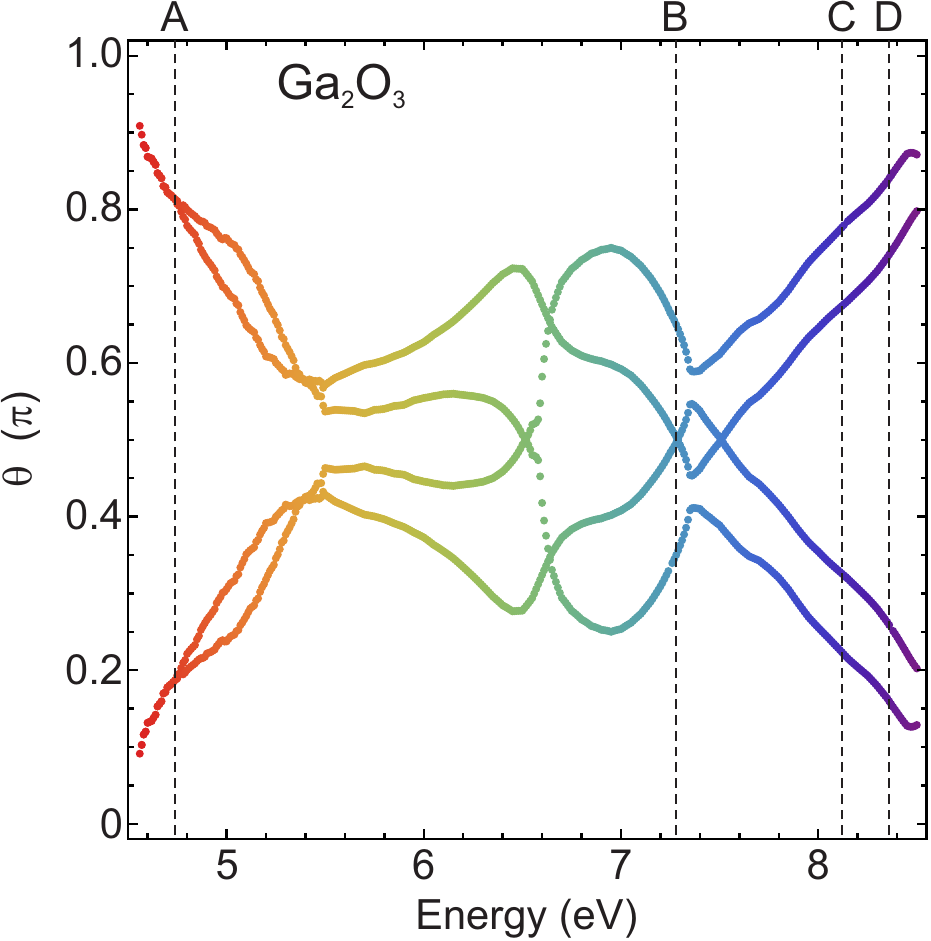}\\
\caption{\label{pts_ga2o3}(a,b) Angular position of the singular optical axes in $\beta$-Ga$_2$O$_3$ as a function of photon energy (a) in the $(\phi,\theta)$-plane (all 8 solutions) and (b) stereographic projection from upper hemisphere (4 solutions)
onto the $(x,y)$-plane with crystallographic directions (marked by open circles) as labelled.
(c) $\phi$ and (d) $\theta$ only as a function of energy. 'A' denotes the beginning of significant splitting of the singular optical axes in the dissipative regime ($E = 4.78$\,eV),
'B' denotes a range ($E \approx 7.23$--$7.33$\,eV) of almost uniaxial degeneracy of the \textit{Windungsachsen} (in the ($x,y$)-plane close to the $x$-direction); 
at 'C' ($E=8.14$\,eV) and 'D' ($E=8.37$\,eV) the material is triaxial, i.e. two of the \textit{Windungsachsen} are exactly degenerate (in the ($x,z$)-plane), causing one 'normal' optical axis and a total of three optical axes}
\end{figure*}

In Fig.~\ref{pts_ga2o3} the spectral dependence of the angular position of the optical axes of $\beta$-Ga$_2$O$_3$ is depicted in detail. Fig.~\ref{pts_ga2o3}a shows the position in the $(\phi,\theta)$-plane and Fig.~\ref{pts_ga2o3}b in stereographic projection. The panels c and d depict the spectral dependence of $\phi$ and $\theta$ alone. 

In the transparency regime there are
two optical axes; they exhibit significant angular splitting into the four singular optical axes at point 'A' at about $E=4.78$\,eV. 

Within the energy range $E \approx 7.23$--$7.33$\,eV marked as 'B', the singular axes are almost (but not quite) oriented along the $[100]$-direction (at $\theta$ close to $\pi/2$ and with $\phi \approx \pi$ and $\phi \approx 0$), rendering the material "almost" uniaxial. However, here the $\phi$-positions do not cross the $\phi=0$ or $\pi$ symmetry line (related to the $(x,z)$ mirror plane of the monoclinic structure). At $E=7.29$\,eV (and at $E=6.52$\,eV and $E=7.50$\,eV), one pair of singular axes is oriented exactly at $\theta=\pi/2$, i.e. within the $(x,y)$-plane (Fig.~\ref{pts_ga2o3}d).

Remarkably at points 'C' ($E \approx 8.14$\,eV) and 'D' ($E \approx 8.37$\,eV), one pair of the singular axes degenerates and lies at $\phi=0$ (or $\pi$) in the $(x,z)$ symmetry plane. Thus at such "triaxial" points, two singular axes remain and one "normal" but absorptive optical axis (without chirality) exists.

\begin{figure*}
\centering
(a)\includegraphics[width=0.27\linewidth]{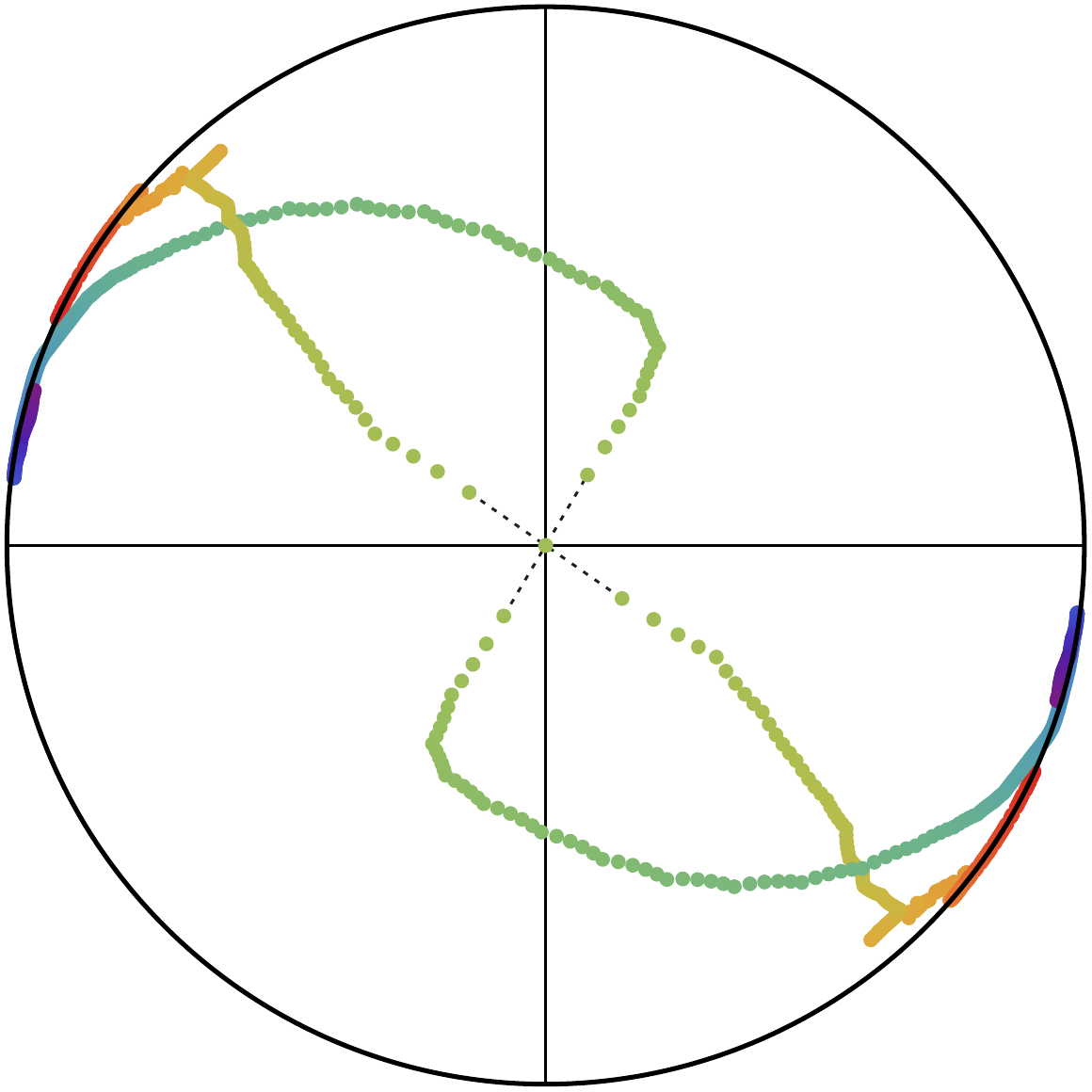}
(b)\includegraphics[width=0.27\linewidth]{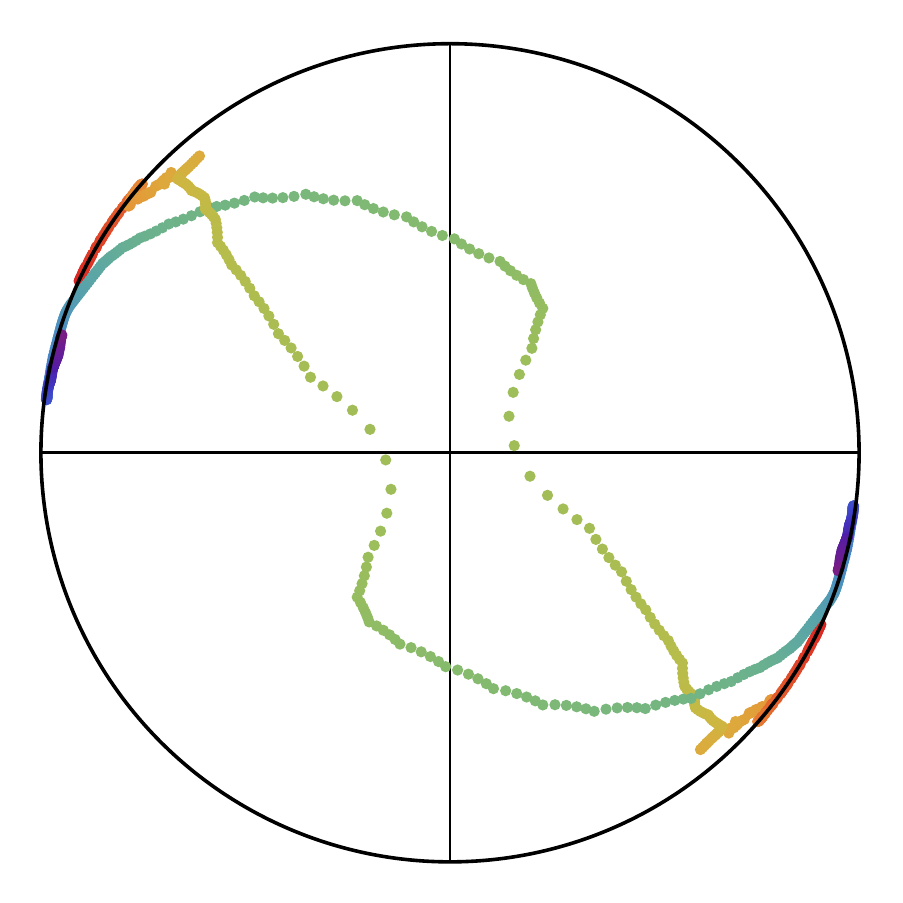}
(c)\includegraphics[width=0.27\linewidth]{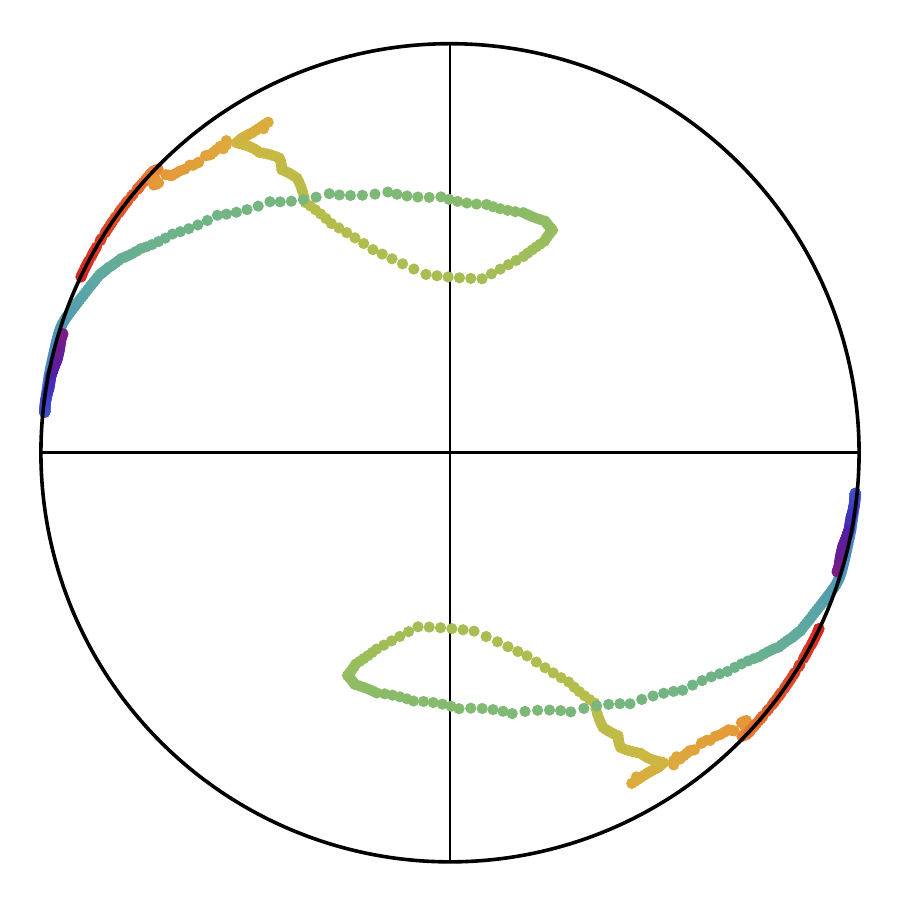} \\
(d)\includegraphics[width=0.27\linewidth]{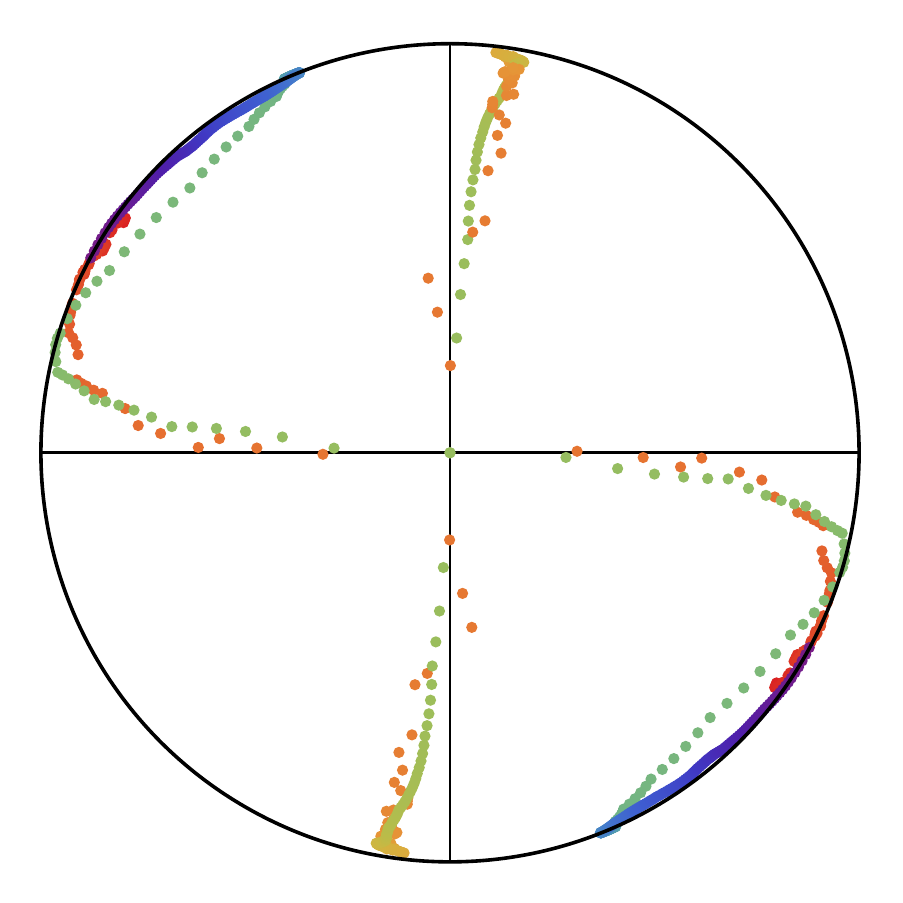}
(e)\includegraphics[width=0.27\linewidth]{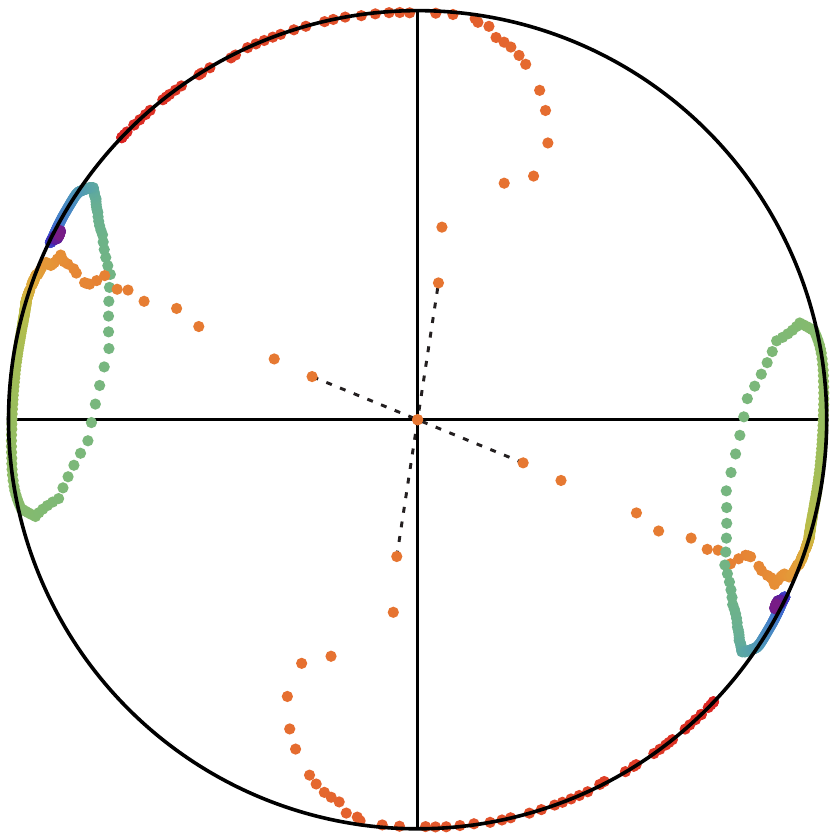}
(f)\includegraphics[width=0.27\linewidth]{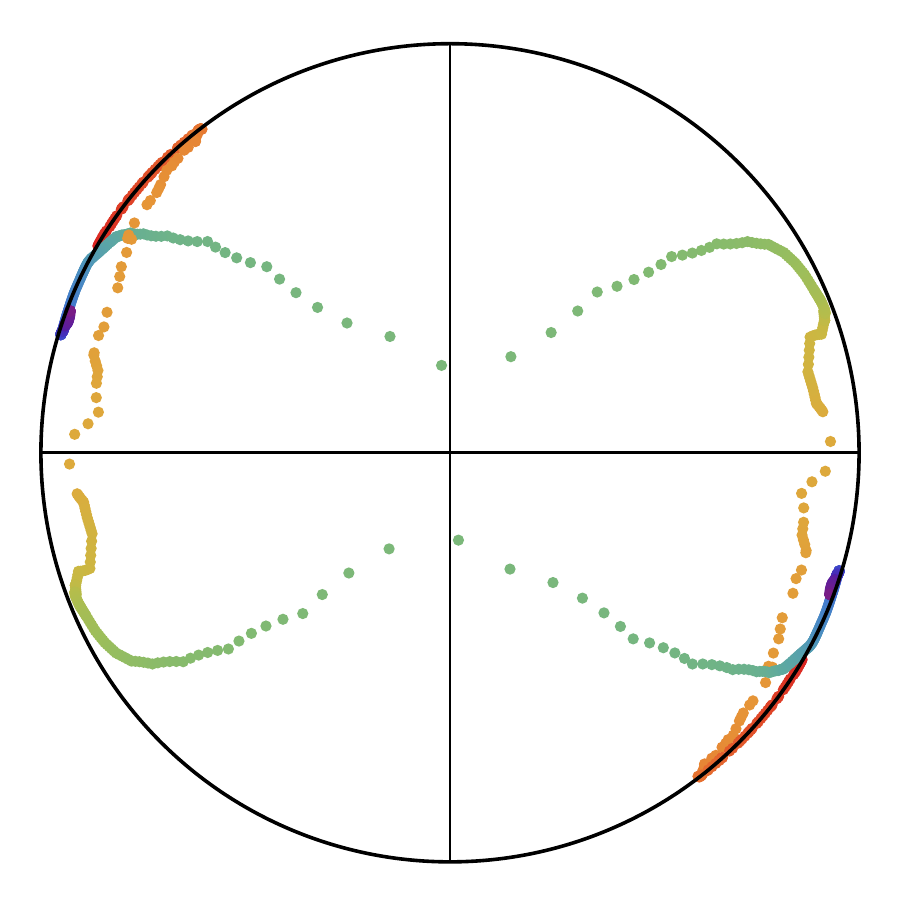}
\caption{\label{fig:stokes} Energy dependence (within 4.5--8.5\,eV in equidistant steps of 0.01\,eV) of Stokes vectors (stereographic projection $(S_1,S_2)$) for (a) the singular optical axis at $E=6.20$\,eV (indicated as 'W' in the icon to Video~\ref{vid:dens}, $\phi=2.170042...$, $\theta=1.754275...$), (b) for a direction close to 'W' ($\phi=2.17$, $\theta=1.75$), exhibiting "anti-crossing" behavior compared to panel (a),
(c) for a direction a little further away from 'W' ($\phi=2.13$, $\theta=1.79$), (d) for the direction where a singular optical axis exists for two different energies ($E \approx 5.02$\,eV and $E \approx 6.29$\,eV, intersection 'X' in Fig.~\ref{pts_ga2o3}b), running twice through $(S_1,S_2)=(0,0)$, i.e. $S_3=\pm 1$, (e) for the direction of the other singular axis at $E \approx 5.02$\,eV ('\,X'\,'), exhibiting only one $S_3=\pm 1$ point, (f) for $\phi=\pi/2$, $\theta=\pi/2$, i.e. the [010]-orientation.}
\end{figure*}

\begin{figure*}[htb!]
\centering
(a)\includegraphics[width=0.44\linewidth]{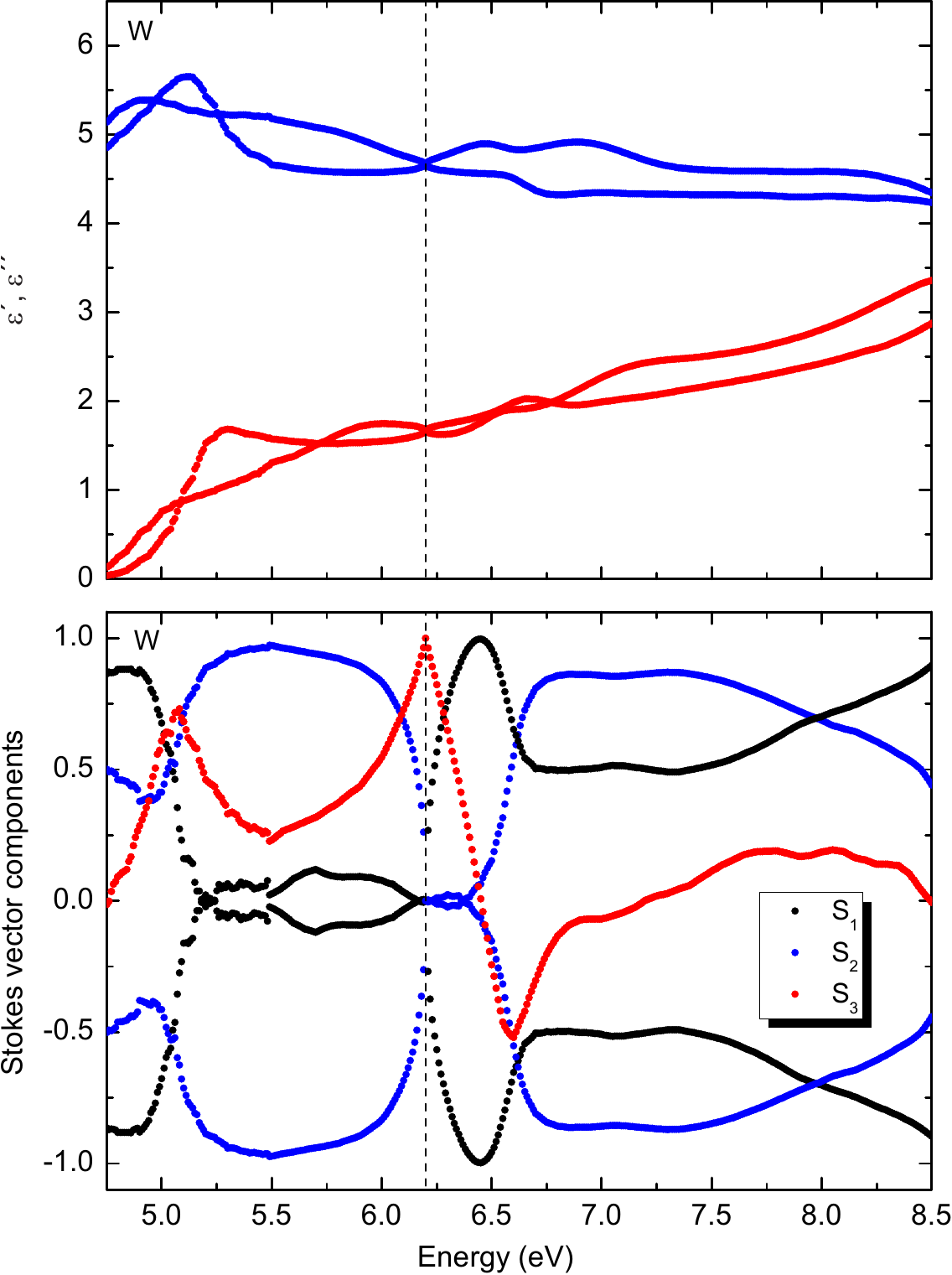}
(b)\includegraphics[width=0.44\linewidth]{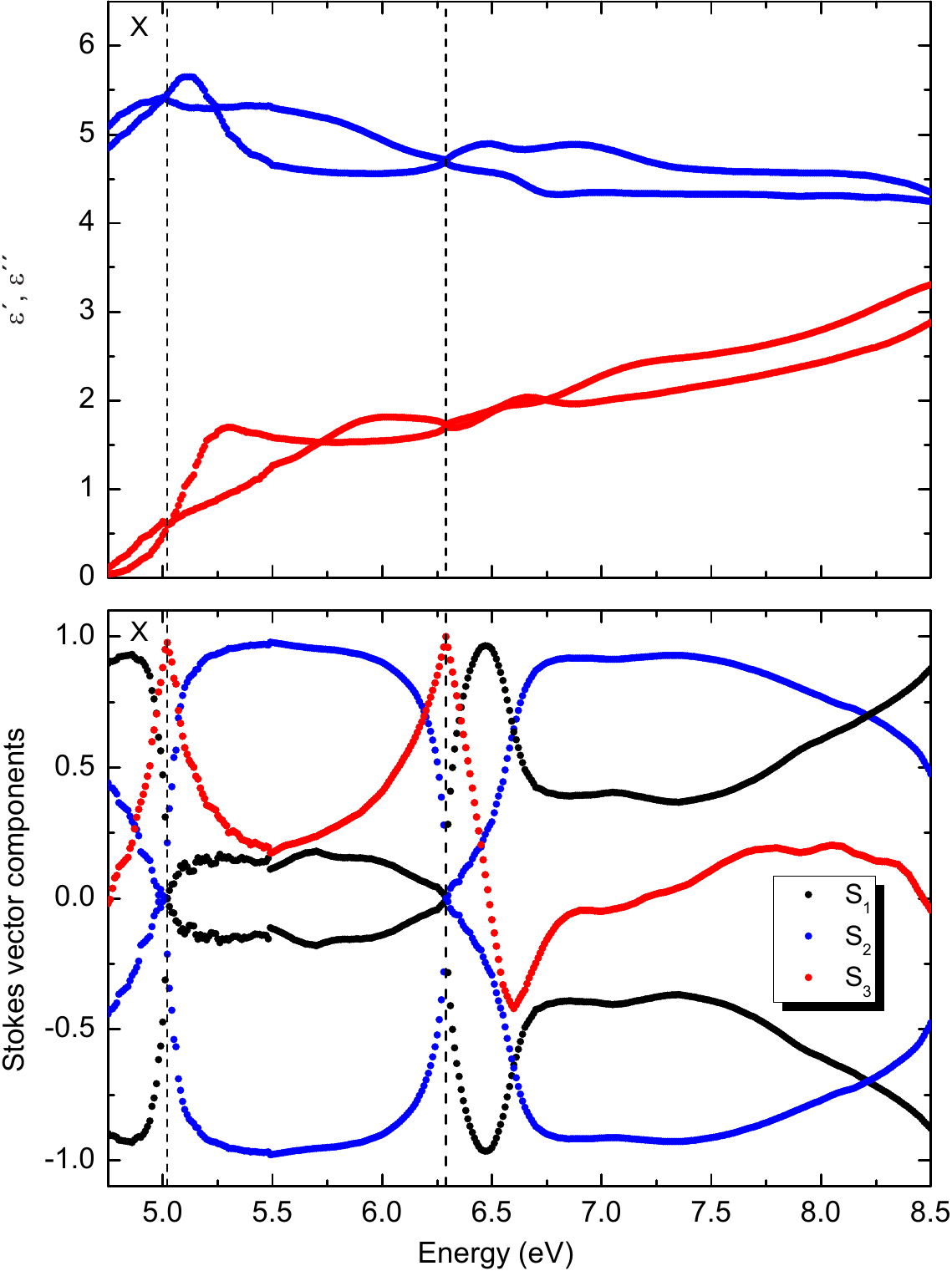}\\
(c)\includegraphics[width=0.44\linewidth]{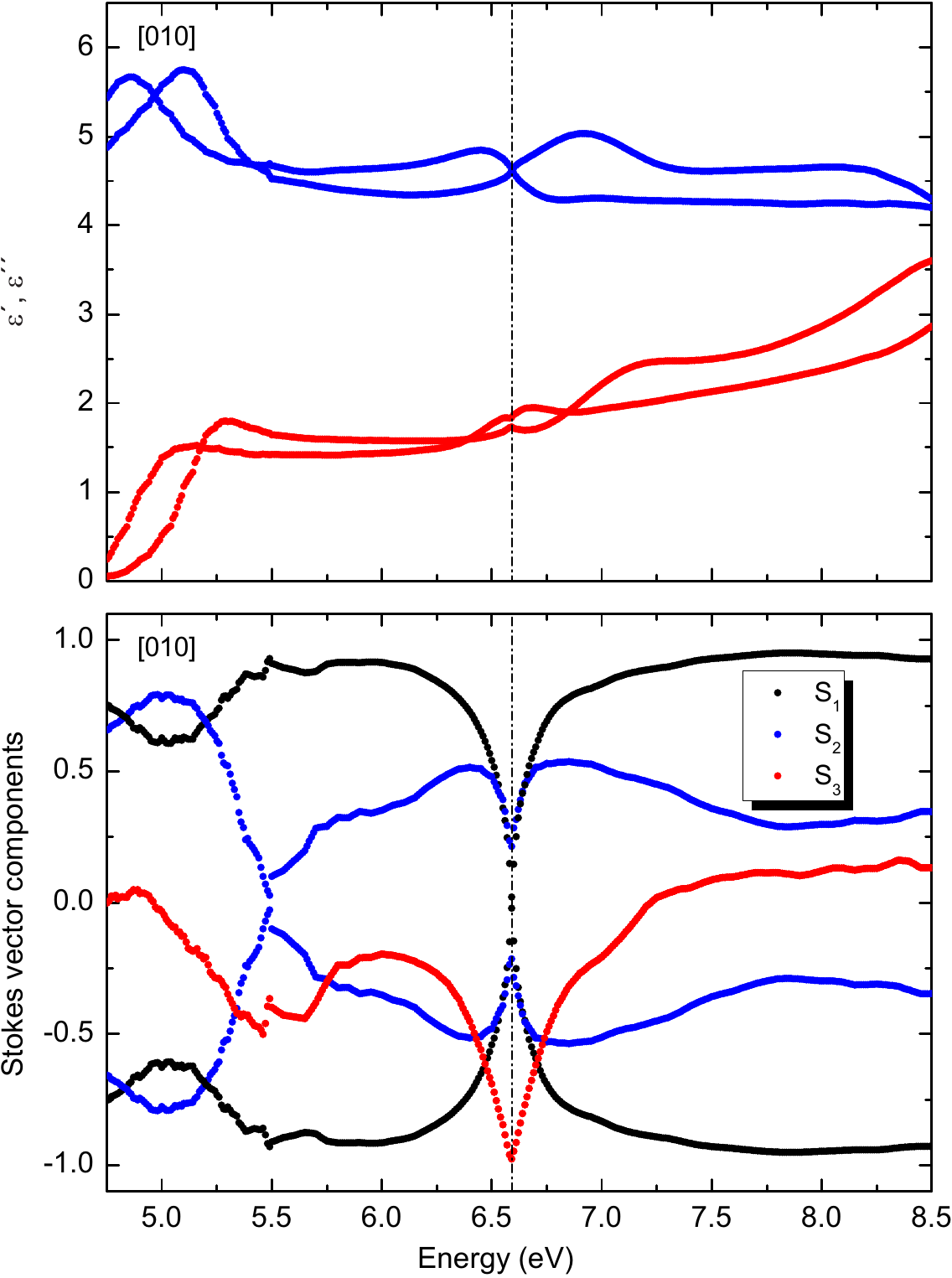}
(d)\includegraphics[width=0.44\linewidth]{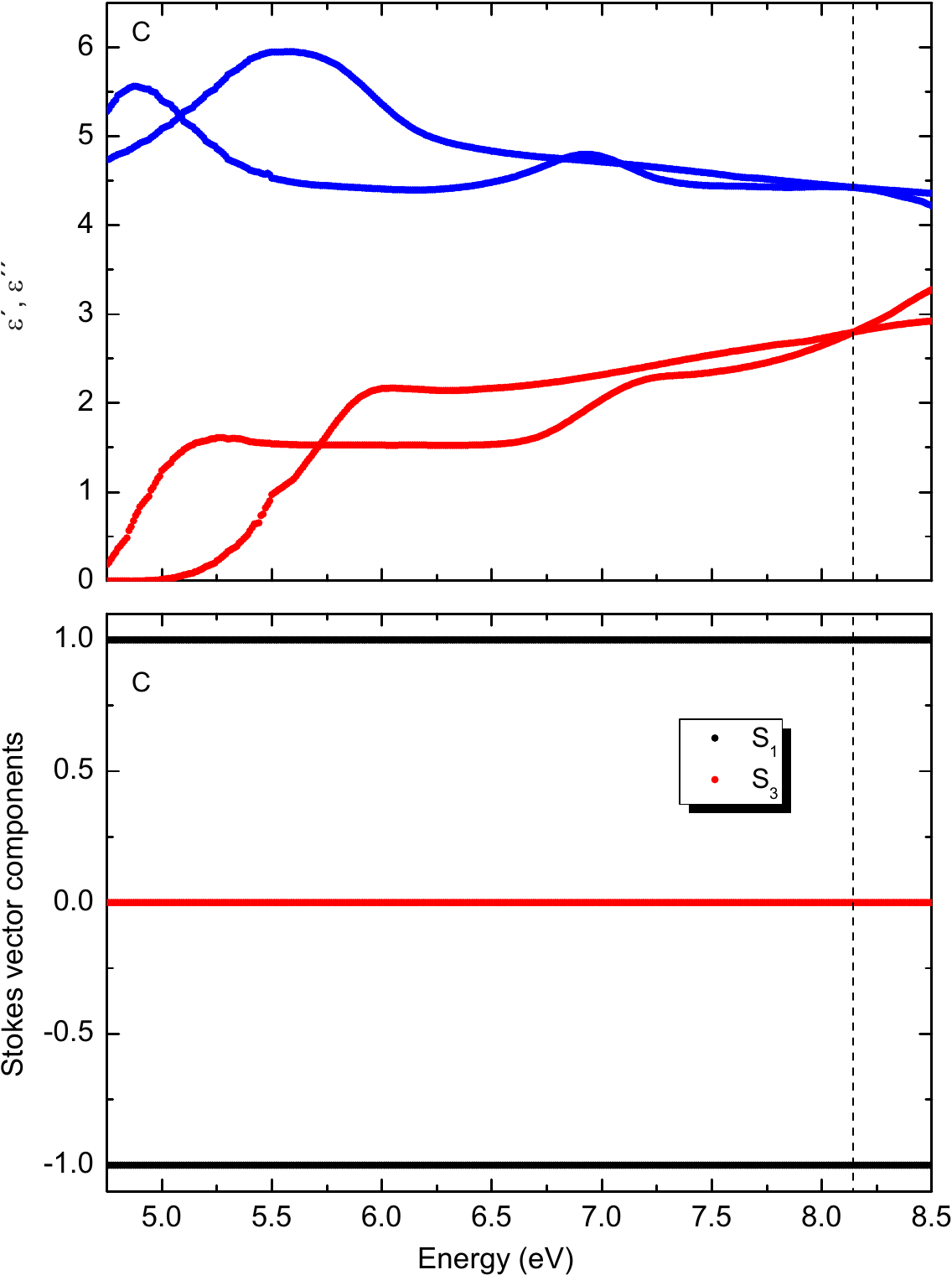}
\caption{\label{fig:all_n2} Energy dependence of the real (\textit{blue}) and imaginary (\textit{red}) parts of $n_1^2$ and $n_2^2$ (upper panels) and the Stokes vector components ($S_1$: \textit{black}, $S_2$: \textit{blue} and $S_3$: \textit{red}) (lower panels). (a) for the direction 'W'. The energy position for which a singular axis exists for this direction is indicted by a \textit{dashed line}. (b) for the direction 'X'. (c) for the $[010]$-direction. The energy position for which "almost" a singular axis exists for this direction ($E=6.59$\,eV, $|S_3|=0.977$) is indicted by a \textit{dashed-dotted line}. (d) for the direction 'C'. The energy position for which a triaxial point and a normal optical axis exists for this direction is indicted by a \textit{dashed line}.}
\end{figure*}

The spectral dependence of the Stokes vectors is visualized for certain directions in Fig.~\ref{fig:stokes} in stereographic projection $(S_1,S_2)$. In Fig.~\ref{fig:stokes}a the Stokes vectors are shown for the orientation 'W' as labelled in the icon of Video~\ref{vid:dens} ($\phi=2.170042...$, 
$\theta=1.754275...$). For $E=6.20$\,eV, the two vectors intersect at $(S_1,S_2)=0$. For deviations from that orientation, the vectors no longer intersect at the pole (Figs.~\ref{fig:stokes}b,c). For the orientation 'W' the Stokes vector components are shown in a different plot in Fig.~\ref{fig:all_n2}a where also the two solutions $n_{1,2}^2=\epsilon=\epsilon' + \imath \, \epsilon''$ are visualized. The singular axis is at that energy where the real \textit{and} imaginary parts of $n_1^2$ and $n_2^2$ are equal. We note that there are other energies where the real \textit{or} imaginary parts of $n_1^2$ and $n_2^2$ are equal.

In Fig.~\ref{fig:stokes}d the Stokes vectors are shown for the orientation 'X' ($\phi=2.246447...$, 
$\theta=0.980297...$) for which a singular axes exists for two different energies, $E \approx 5.02$\,eV and $E \approx 6.29$\,eV; accordingly the Stokes vectors intersect at $S_3=1$ twice (see also Fig.~\ref{fig:all_n2}b). In Fig.~\ref{fig:stokes}e the Stokes vector for the other orientation of the singular optical axis at $E=5.02$\,eV is shown that of course displays a $S_3=\pm 1$ Stokes vector only once in the displayed energy range. 

Finally in Fig.~\ref{fig:stokes}f the Stokes vectors are shown for the $[010]$-orientation ($\phi=\pi/2$, $\theta=\pi/2$). At $E=6.59$\,eV a fairly large circularly polarized component ($S_3=-0.977$) exists, but in the investigated spectral range there is no singular axis oriented exactly in this direction. The according solutions $n_1^2$ and $n_2^2$ and the Stokes vector components for the $[010]$-orientation are shown in Fig.~\ref{fig:all_n2}c.

For the orientation 'C' ($\phi=0$, $\theta=0.682068...$) the solutions $n_1^2$ and $n_2^2$ and the Stokes vector components are shown in Fig.~\ref{fig:all_n2}d. Since this point is within the $(x,z)$-plane the eigen-polarizations are always along $x$ and $y$ ($S_1=\pm 1$) and $S_2 \equiv 0$ and $S_3 \equiv 0$ except for the energy of the triaxial point when for the present "normal" optical axis the Stokes vector is undefined.

\section{Summary and Outlook}

In summary we have categorized biaxial crystals with regard to the degeneracy of the index of
refraction of their four singular optical axes in the absorption regime. Triclinic, monoclinic 
and orthorhombic crystals have in general 4, 2 and one different complex index of refraction, respectively.
Thus the different crystal symmetries leads to distinguishable optical properties. We note that the 
three crystal systems that lead to optically uniaxial crystals (tetragonal, trigonal and hexagonal)
cannot be distinguished optically.
Analytical formula have been given for the complete angular positions of the singular axes
in the orthorhombic case, $(\phi,\theta)(\epsilon)$, and
at least for their azimuthal position in the monoclinic case, $\phi(\epsilon)$.

The spectral dispersion of the singular optical axes in a mono\-clinic material ($\beta$-Ga$_2$O$_3$) has been shown
and accidental degeneracies deliver an "almost" uniaxial crystal (all four singular axes are very close) 
in a certain spectral range.
At two distinguished spectral positions the crystal is found to be triaxial when two of the four singular axis are exactly degenerate, creating the novel case of a "normal" but absorptive optical axis and leaving two singular optical axes.

We think that the analysis scheme presented here will be useful to model, predict and analyze in detail and properly
the optical properties
of biaxial materials in photonic devices operating in the absorption and gain regime.

\begin{acknowledgments}
We acknowledge an encouraging discussion with Sir Michael Berry and fruitful exchange with R\"udiger Schmidt-Grund. 
This work has been funded by Deutsche Forschungsgemeinschaft in the framework of Sonderforschungsbereich SFB 762 "Functionality
of Oxide Interfaces" and supported by Forschungsprofilbereich "Complex Matter" of Universit\"at Leipzig.
\end{acknowledgments}

\end{document}